\documentclass[pre,10pt,twocolumn]{revtex4-1}
\usepackage[utf8]{inputenc}
\usepackage{graphicx}
\usepackage{amsmath}
\usepackage{amssymb}
\usepackage{textcomp}
\graphicspath{{figures/}}

\begin{document}

\title{Efficient manifolds tracing for planar maps}
\author{D. Ciro, I.L. Caldas, R.L. Viana and T.E. Evans}
\date{\today}

\begin{abstract}
In this work we introduce an exact calculation method and an approximation technique for tracing the invariant manifolds of unstable periodic orbits of planar maps. The exact method relies in an adaptive collocation procedure that prevents redundant calculations occurring in non-refinement approaches and the approximated method is based on an intuitive geometrical decomposition of the manifold in bare and fine details. The resulting approximated manifold is precise when compared to the exact manifold, and its calculation is computationally more efficient, making it ideal for mappings involving intensive calculations like numerical function inversion or the numerical integration of ODEs between crossings through a surface of section.
\end{abstract}
\maketitle

\section{Introduction}
In dynamical systems, the determination of unstable periodic orbits and their invariant manifolds is a problem on its own, and is fundamental to understanding the underlying structure of the chaotic orbits of planar maps~\cite{ozorio1988}. These manifolds provide the skeleton for chaotic dynamics and are relevant in physical applications for the interpretation of the transport induced by non-integrability~\cite{litchemberg1992, meiss1992}.

Physically, invariant manifolds determine the geometrical features of advection and delimit the extension of irregular flows. For instance, in magnetically confined plasmas a small non-axisymmetric magnetic field induces Hamiltonian chaos near the plasma edge of single-null tokamak discharges~\cite{da-silva2002, evans2002, morrison2000, wingen2007, wingen2014}. In this situation, the invariant manifolds of the magnetic saddle can be related to the heat flux patterns where the confinement chamber interacts with the plasma~\cite{evans2005, ciro2017}. In analogous systems, time-dependent perturbations lead to orbit transfers between otherwise isolated regions of bistable non-linear oscillators~\cite{reichl1984}. Here invariant manifolds determine the transport channels between the phase space domains.

From the dynamics point of view, manifold tracing is required for determining of the geometry of the basins of attraction of chaotic sets~\cite{aguirre2009}, and can be used to identify homoclinic and heteroclinic intersections which are contained in non-attracting invariant sets that play a fundamental role on the description of the dynamics of typical chaotic orbits in the phase space~\cite{guckenheimer1983}. Although global quantities like the Liapunov exponents allow to determine numerically the bifurcation parameters, they do not, in general, provide details on the type of bifurcation that takes place. In this situations, precise invariant manifolds tracing determines clearly the topological mechanisms that take place during the bifurcations and accounts for physically relevant features, like regions of enhanced transport or the destruction of transport barriers~\cite{kroetz2008}.

In some situations, the geometry of the manifolds can be estimated through the mapping of a large collection of orbits close to the saddle, a procedure similar to the used to determine chaotic saddles~\cite{kantz1985, nusse1989, moresco1999, sweet2001}. However, without an ordering scheme and refinement this method is computationally expensive and limited in resolution.

In this work we introduce an \emph{exact} and an approximation method to calculate the invariant manifolds of periodic saddles of planar maps. The maps in question can be explicit or induced by three-dimensional flows. Both calculations are based in the manifold decomposition in primary segments~\cite{hobson1993}. The \emph{exact} method relies on the efficient calculation of primary segments from a single seed segment near the saddle, while the approximated method is based on the determination of reliable interpolants for arbitrary primary segments. The latter is specially relevant for maps induced by flows, where lengthly numerical integrations are required to produce a single iteration of the Poincaré map. This allow us to deal with the increase in length of the manifold segments due to the stretching and folding mechanism inherent to the chaotic dynamics.

The approximation method involving the primary segments interpolation has been discussed before by Hobson in~\cite{hobson1993}, and more detailed implementations were made by Goodman and Wróbel in \cite{goodman2011} and references therein. In this work  we introduce a versatile interpolation approach based in the curve decomposition in bare and fine details, which allow us to focus on different aspects of the manifold separately. The bare details of the curve are determined by a suitable discretization procedure that determines an appropriate set of nodes containing most of the curve information, while the fine details are contained in a normal displacement function or \emph{shape function} which ensures the smoothness of the manifold. The presented normal displacement description is new to our knowledge and can be easily implemented, containing a single set of adjustable parameters which can be fixed from intuitive geometrical conditions. Complementing other approaches based in Catmull-Rom splines~\cite{goodman2011}. The resulting approximation method gives the invariant manifold as a continuous parametric curve from which we can measure the distance to the discretized manifold obtained from the exact calculation. This allows to estimate the approximation error that, in general, is well below an estimated baseline.  The approximation procedure is stable and more efficient that the exact one, making it ideal for precise and fast calculations.

This work is not intended to compare existing methods in the literature with the introduced approaches, but to enable the reader to trace exact and approximated planar manifolds with a reliable procedure for tracing invariant manifolds, without dipping into the subject of numerical interpolation. The paper is organized as follows, in Sect.~\ref{sec.primary_segments} we present a review on invariant manifolds and their representation through primary segments. In Sect.~\ref{sec.mapping_refinement} we present an \emph{exact} calculation method and in in Sect.~\ref{sec.interpolant_mapping} we introduce its approximated version, where the interpolation procedure is detailed. In Sect.~\ref{sec.example_application}, we show an example application of the manifold tracing for bi-stable systems where the transition from homoclinic to heteroclinic chaos is used to explain the transition from alternating chaos to global chaos. In Sect.~\ref{sec.conclusions} we present our conclusions and perspectives.

\section{Invariant manifolds and primary segments}\label{sec.primary_segments}
In this section we want to introduce the concept of invariant manifolds for planar maps and show their decomposition in primary segments. We also show, formally, the method to obtain any primary segment from a single known segment and illustrate this with the first order parametric form of the manifold near the periodic orbit.

\begin{figure}[h]
 \centering
 \includegraphics{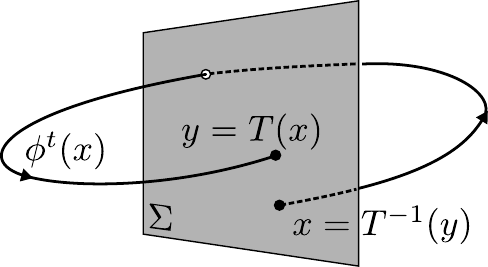}
 \caption{\label{fig.poincare_map} Poincaré map as the first return to the surface $\Sigma$.}
\end{figure}

Without loss of generality we will develop our discussion around the Poincar\'e maps of three-dimensional flows, since they represent a situation in which we do not have access to a closed-form expression for the planar map of interest. Consider the dynamical system
\begin{equation}
 \frac{d x}{dt} = f(x),
\end{equation}
where $x\in \mathbb{R}^3$ and $f: \mathbb{R}^3\rightarrow\mathbb{R}^3$. The solution $\phi^t(x_0)$, of this differential equation takes an initial condition $x_0\in\mathbb{R}^3$ and maps it to a new point $x\in\mathbb{R}^3$ for a given time $t\in\mathbb{R}$. Now, consider an orientable two-dimensional surface $\Sigma$ which is transverse to the flow. Then, the Poincaré map $T$, is a diffeomorphism $T: \Sigma\rightarrow\Sigma$, such that $y = T(x)$, where $x,y\in\Sigma$ and $y = \phi^\tau(x)$ for the smallest $\tau>0$, such that the flow traverses $\Sigma$ always in the same direction (Fig.~\ref{fig.poincare_map}). Analogously, the \emph{inverse map} $T^{-1}$ is such that $x = T^{-1}(y)$, where $x = \phi^\tau(y)$ for the largest $\tau<0$.

The fixed point $x^*\in\Sigma$, satisfies $T(x^*) = x^*$, which means that the orbit starting in $x^*$ is closed or periodic. For the case of interest $x^*$ is a saddle point, i.e. the Jacobian matrix of $T$ has real eigenvalues $\{\lambda_s, \lambda_u\}$, such that $|\lambda_s|<1$ and $|\lambda_u| > 1$. 

The unstable manifold of $x^*$, $\mathcal{W}_u(x^*)$, is the collection of points that converge to $x^*$ under $T^{-1}$, and the stable one $\mathcal{W}_s(x^*)$ is the collection that converge to $x^*$ under $T$~\cite{guckenheimer1983}.
\begin{eqnarray}
 \mathcal{W}_u(x^*) &=& \{x\in\Sigma: T^{-n}(x)\rightarrow x^*\mbox{, as } n\rightarrow\infty\}\\
 \mathcal{W}_s(x^*) &=& \{x\in\Sigma: T^n(x)\rightarrow x^*\mbox{, as } n\rightarrow\infty\}
\end{eqnarray}

\begin{figure}[h]
 \centering
 \includegraphics{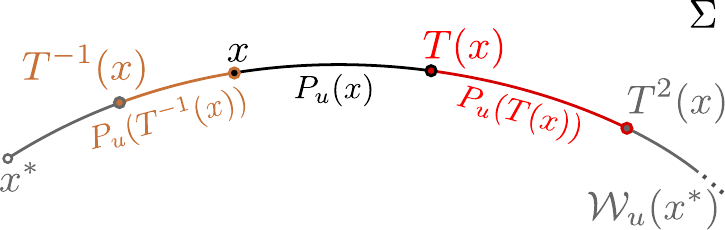}
 \caption{\label{fig.primary_segment} The unstable manifold can be subdivided as a sequence of primary segments based on the images and pre-images of an arbitrary point of the manifold}
\end{figure}

For the numerical calculation of the invariant manifolds it is useful to develop an ordering scheme that allow us to calculate them recursively, for this we resort to the concept of \emph{primary segment} introduced in \cite{hobson1993}. A primary segment based on $x$, is a continuous subset of the manifold $P_{s,u}(x)\subset\mathcal{W}_{s,u}(x^*)$, containing every point in $\mathcal{W}_{s,u}(x^*)$ between $x$ and $T(x)$, including $x$ but not $T(x)$. Notice that, if $y\in P_u(x)$, then $T(y)\notin P_u(x)$, but, clearly $T(y)\in P_u(T(x))$, i.e. every point inside $P_u(x)$ has an image in the next primary segment and a pre-image in the previous one, and this is also valid for the base point $x$ (Fig.~\ref{fig.primary_segment}).
Without loss of generality we will continue this discussion in terms of the unstable manifold, since the definitions and methods are analogous for the stable one.

The unstable manifold can be written as the union of all primary segments based on the images of an arbitrary $x\in\mathcal{W}_u(x^*)$
\begin{equation}
 \mathcal{W}_{u}(x^*) = \bigcup_{n=-\infty}^{\infty} P_u(T^n(x)).
\end{equation}
Since every point in $P_u(x)$ has an image in $P_u(T(x))$ we can say that the primary segment based on $T(x)$ is given by
\begin{equation}
 P_u(T(x)) = T[P_u(x)],
\end{equation}
where $T[A]$ is the set obtained after mapping every element of $A$. This implies that the manifold can be obtained by joining subsequent mappings of an arbitrary primary segment
\begin{equation}
 \mathcal{W}_{u}(x^*) = \bigcup_{n=-\infty}^{\infty} T^n[P_u(x)].
\end{equation}
Consequently, if we know every point in any primary segment, we can map them forward and backwards repeatedly to obtain the desired extent of the manifold. This is the theoretical basis of interpolation methods~\cite{hobson1993, goodman2011}. Now, we need to determine at least one primary segment of the manifold. This can be done approximately in the region near the saddle point. Consider an arbitrary point $x_\epsilon$ close the fixed point $x^*$, whose image $T(x_\epsilon)$ is also close to $x^*$, then to a first order
\begin{equation}\label{eq.first_order_Tx}
 T(x_\epsilon) \approx x^* + J_T(x^*)(x_\epsilon-x^*)
\end{equation}
where $J_T$ is the Jacobian of the Poincaré map. Now, assume that $x-x^*$ is parallel to the eigenvector of $J_T$ with eigenvalue $|\lambda_u| > 1$, i.e. $x_\epsilon = x^* + \epsilon\vec v_u$, with $|\vec v_u| = 1$ and $\epsilon<<1$, so that the mapping of $x_\epsilon$ becomes
\begin{equation}
 T(x_\epsilon) \approx x^* + \epsilon\lambda_u\vec v_u.
\end{equation}
Consequently, the \emph{zero}'th primary segment can be approximated by the straight segment joining $x$ and $T(x)$.
\begin{equation}\label{eq.first_primary}
 P_0(x_\epsilon) = \{ L_\epsilon (t)  \mbox{, } 0\leq t <1\},
\end{equation}
where
\begin{equation}\label{eq.param_line}
 L_\epsilon (t) = (1-t)(x^*+\epsilon\vec v_u) + tT(x^*+\epsilon\vec v_u),
\end{equation}
provided that $\epsilon$ is sufficiently small. In regular situations with $\epsilon<<1$, the first order parametric form of the primary segment is sufficient, but particular situations might require higher order terms in the expansion of the local manifold~\cite{fuming1994}, also, the map $T$ must be replaced by $T^2$ if the eigenvalue $\lambda_u$ is negative.

The unstable manifold can then be written as the following semi-infinite union
\begin{equation}\label{eq.unst_man_union}
 \mathcal{W}_u(x^*, T) = \widetilde{x^*x_\epsilon}\cup P_0\cup P_1\cup\cdots\cup P_i\cdots
\end{equation}
where $P_{i+1} = T[P_{i}]$ and $\widetilde{x^*x}$ is the union of inverse mappings of $P_0$ which can be approximated by the straight segment $\overline{x^*x_\epsilon}$.

With this simple definition we can compute a finite representation of the invariant manifold to the desired extension, provided we know how to calculate $T[P_i]$ and we know one segment $P_0$. Formally speaking, if we know how to compute $T(y)$ for any $y\in P_i$, then we are able to compute $T[P_i]$, but mapping a set of uniformly distributed points in $P_i$ will not, in general, lead to evenly distributed set of points in $P_{i+1}$. 

As we do not know \emph{a-priori} which distribution of points in $P_0$ will lead to a uniform discretization of $P_n$, an interim solution consists in discretizing $P_0$ uniformly with a large number of nodes, although, eventually, the stretching and folding mechanism will overcome the excess points, and the segments will become poorly resolved.
\begin{figure}[h]
 \centering
 \includegraphics[]{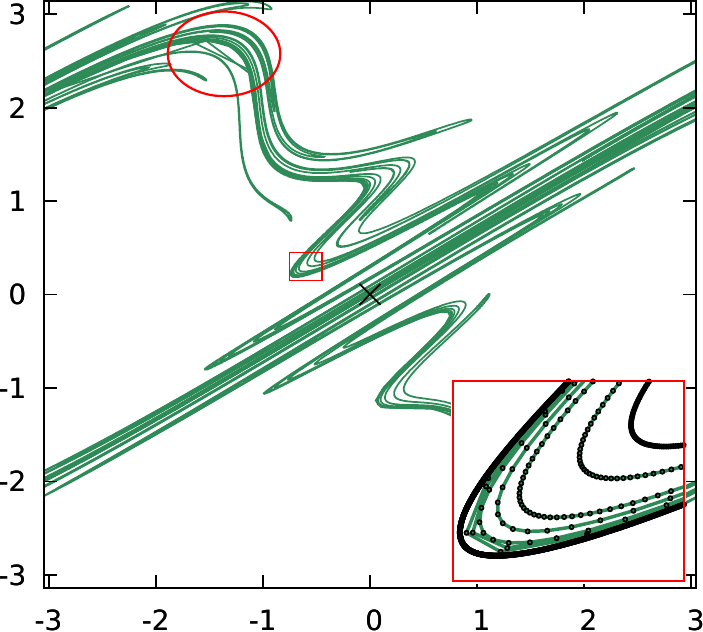}
 \caption{\label{fig.unst_man_st_map_simple} Unstable manifold portion of the saddle $x^*=(0,0)$ for the Chirikov-Taylor map with $k=1.5$. The base segment $P_0$ was uniformly discretized with $4662$ points, and the portion contains $25$ primary segments. In the red circle we can see a poorly resolved region and the red inset shows the effect of stretching for different layers of a lobe.}
\end{figure}

In Fig.~\ref{fig.unst_man_st_map_simple} we show a short portion of the unstable manifold of the period-\emph{one} saddle of the Chirikov-Taylor map~\cite{chirikov1979},
\begin{eqnarray}
 y_{n+1} &=&  y_n+k\sin x_n, \\
 x_{n+1} &=& x_n+y_{n+1},
\end{eqnarray}
for $k=1.5$. Although the resulting curve appears smooth, close inspection shows the undesired effects of uniform discretization in $P_0$. The segment stretching leads to a complex mix of over-resolved and under-resolved portions of the manifold. In Section~\ref{sec.mapping_refinement}, we introduce an adaptive method that allow us to determine the discretization of $P_0$ that leads to a well resolved discretization of an arbitrary $P_n$.

\section{Mapping-Refinement method}\label{sec.mapping_refinement}
In this section we introduce a refinement method for producing a suitable discretization of the~\emph{exact} manifold which satisfies some predefined resolution criteria that limits the distance between subsequent points and the angles between subsequent secant lines. The resulting set of nodes can be regarded as a representation of the exact manifold, although to be precise we would need to know the curve between each pair of nodes, and this can not be exactly achieved in the practice.

Consider the following initial discretization of $P_0$
\begin{equation}
 P_0 \rightarrow \{x_{0,1}, x_{0,2},...,x_{0,M}\},
\end{equation}
where the '$\rightarrow$' symbol is used instead of '$=$' to indicate that we are only \emph{representing} the continuous set $P_0$ by a discrete set of points. The initial set of points can be obtained by evaluating 
\begin{equation}
x_{0,i} = L_\epsilon \left(\frac{i-1}{M}\right),
\end{equation}
for $i=1, 2,...,M$, and $M$ is a small number, usually $<10$. Then the induced discretization of the segment $P_n$ is
\begin{equation}
 P_n \rightarrow \{x_{n,1}, x_{n,2},...,x_{n,M}\},
\end{equation}
where
\begin{equation}\label{eq.point_of_Tn}
 x_{n,i} = T^n(x_{0,i}).
\end{equation}

\begin{figure}[h]
 \centering
 \includegraphics[]{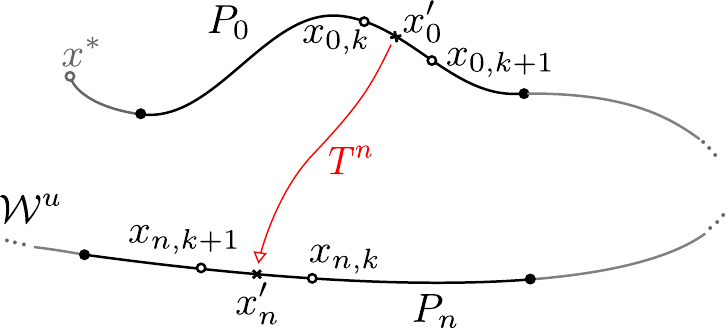}
 \caption{\label{fig.refinement_exact} The discretization of the $n$'th segment may be improved by inserting new points at specific locations in the \emph{zero}'th segment and applying $T^n$ to populate the $n$'th segment in a controlled fashion.}
\end{figure}

As mentioned before, the resulting points of $P_n$ will be unevenly spaced due to the non-uniform stretching of the primary segments, but this can be solved by adding new points to $P_n$ at specific locations. For instance, if some resolution criterion is violated between $x_{n,k}$ and $x_{n,k+1}$, we need to insert a new manifold point $x_n'$ between these points. Since we do not have a parametric representation for $P_n$ we need to obtain the new points from initial conditions in the first segment $P_0$, where we do have a parametric form given by (\ref{eq.first_primary}) (or a higher order version of it). Then we calculate $x_0'$ between $x_{0,k}$ and $x_{0,k+1}$ in $P_0$ and calculate $x_{n}' = T^n(x_0')$, which is the required refinement point (Fig.~\ref{fig.refinement_exact}). This procedure can be iterated until the discretization of $P_n$ satisfies the desired resolution requirements.

The refinement procedure must result in evenly spaced points for low curvature regions and a higher density of points in high curvature regions. Such scale-invariant resolution is desired if we require a reliable description of densely packed portions of the curve, or if we need to study the intersection between manifolds.
\begin{figure}[h]
 \centering
 \includegraphics[]{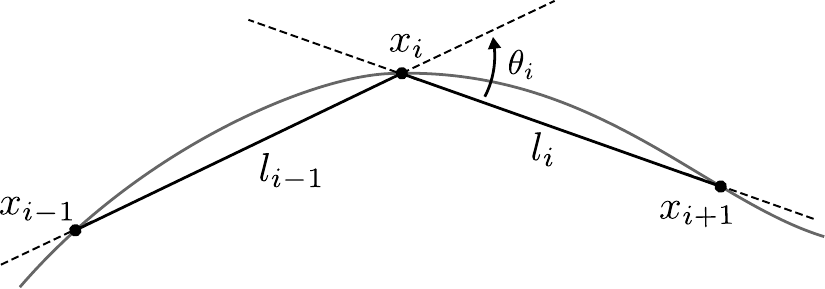}
 \caption{\label{fig.criterion_geometry} Relevant quantities for the resolution criteria. The chord $l_i$ connects the nodes $x_i$ and $x_{i+1}$ and the angle $\theta_i$ is defined between the secant lines containing $l_i$ and $l_{i-1}$.}
\end{figure}
In Fig.~\ref{fig.criterion_geometry} we show the quantities involved in resolution conditions. In general, the length of the chord $l_i$ connecting the nodes $x_i$ and $x_{i+1}$ is required to be below some small value $l_c$. This chord belongs to an infinite secant line that forms and angle $\theta_i$ with the secant line containing the chord $l_{i-1}$. Analogously, the inter-secant angle $\theta_i$ must be below some predefined value $\theta_c$. Provided that $l_c$ and $\theta_c$ are sufficiently small, the arcs connecting the points $x_{i-1}, x_i$ and $x_{i+1}$ will not differ much from straight lines. With this, we establish that most of the geometric information is contained in the nodes instead of the arcs connecting them. Then, the small details contained in the parameters that determine the arcs are used to guarantee the manifold differentiability.

\begin{figure}[h]
 \centering
 \includegraphics[]{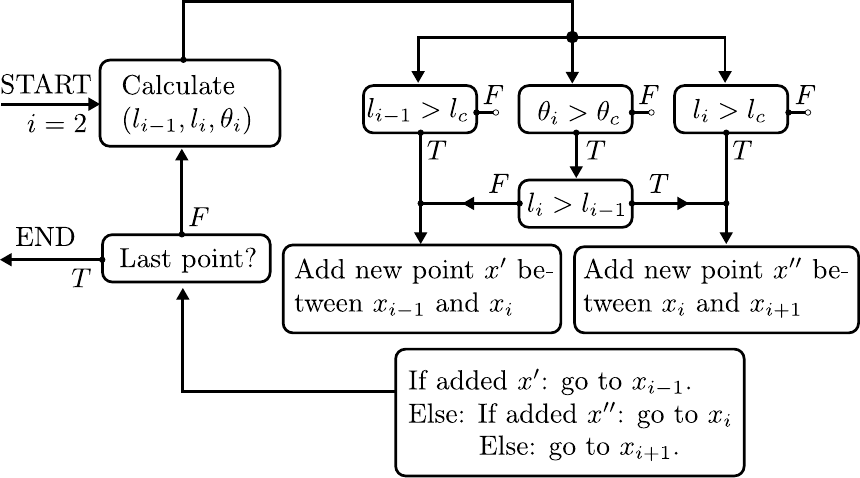}
 \caption{\label{fig.refinement_procedure} Refinement procedure to produce a discretization of $P_n$ satisfying the resolution criteria and preventing superfluous clustering. The points $x'$ and/or $x''$, are obtained by applying $T^n$ to corresponding points in $P_0$.}
\end{figure}

Upon violation of the resolution criteria a refinement procedure must be carried out in an specific order to prevent clustering of new points. In Fig.~\ref{fig.refinement_procedure} we summarize the main steps of the refinement method. In short, a new point is introduced between $x_i$ and $x_{i+1}$ if the chord $l_i$ is larger that the critic distance $l_c$, or if $l_i$ is the longest chord in an angle that exceeds a small critical value $\theta_c$. The procedure is iterated until the criteria are satisfied for all nodes in the segment $P_n$, then these nodes are mapped to obtain an initial discretization of $P_{n+1}$ and so on.

\begin{figure}[h]
 \centering
 \includegraphics[width=0.45\textwidth]{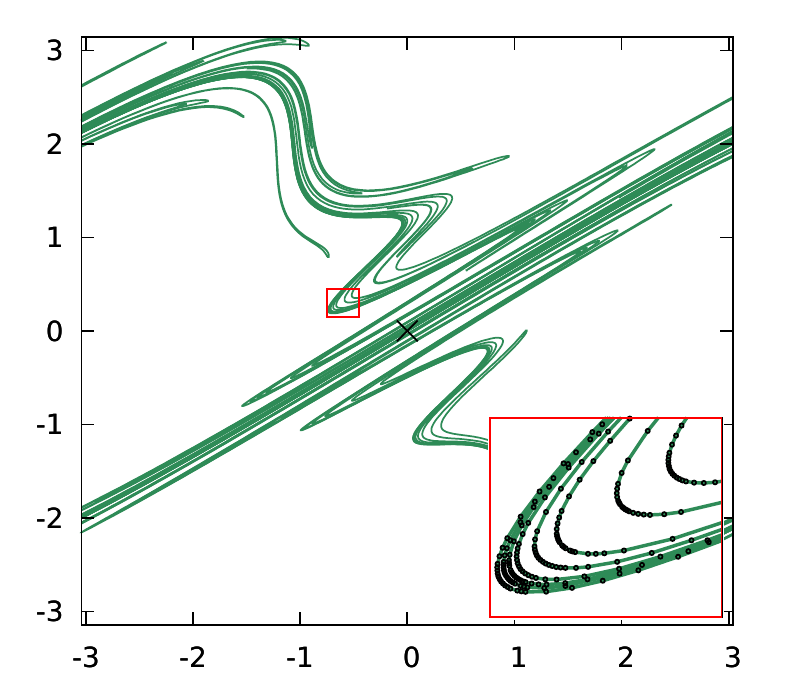}
 \caption{\label{fig.unst_man_st_map} Unstable manifold portion of the Chirikov-Taylor map saddle with $k=1.5$. The adaptive generation of initial conditions on segment $P_0$ was used to satisfy $l_c=0.1$, {$\theta_c=10$\textdegree} for each of the 25 segments depicted.}
\end{figure}

In Fig.~\ref{fig.unst_man_st_map} we show the same portion of Fig.~\ref{fig.unst_man_st_map_simple} calculated with the mapping-refinement method. The calculation was done so that the number of nodes in $P_0$ is the same for the uniform discretization example, and the number of calls to the map function is the same for both calculations. In the inset region of Fig.~\ref{fig.unst_man_st_map} it is clear that the resulting nodes are evenly spaced in regions of low curvature and accumulate in the high curvature regions as required. Although the results from the mapping-refinement method can be regarded as \emph{exact}, they have some inherent error sources, as the initial conditions rely in a continuous representation of $P_0$, which is an approximation, and on the numerical integration routine when we use a Poincar\'e map. Due to the robustness of this method, it is preferred over others, but we can introduce a number of approximations that reduce substantially the computational cost, enabling us to calculate longer extensions of the invariant manifolds.

In the following we discuss a relevant limitation of the presented method before introducing the interpolation-mapping approach. Since all the points in the manifold are seeded in $P_0$, the mapping-refinement method is numerically intensive, more precisely if $M_n$ is the number of points required to resolve the $n$'th segment, the total number of calls to the map function when tracing $n$ segments of a manifold is
\begin{equation}
 Nc(n) = n M_n.
\end{equation}
Assuming that the stretching mechanism leads on average to an exponential growth in the primary segments we obtain
\begin{equation}
 Nc(n) \propto n e^{\lambda n},
\end{equation}
for an appropriate $\lambda > 0$. Consequently, the presented method can be computationally expensive for manifold calculations involving a large number of primary segments, although in most situations a few tens of segments are sufficient to obtain sufficient information on the manifold geometry. A more fundamental problem lies in the finite-precision of the numerical representation, which limits the number of different initial conditions that can be represented in $P_0$, and can be insufficient to resolve a given $P_n$ due to the exponential growth of $M_n$. In practical situations the stretching mechanism can be sufficiently large, for two neighbor initial conditions in $P_0$ separated by the minimum representable numerical difference in coordinates to become separated after a few mappings by a distance that exceeds $l_c$ in the phase space. When this occurs we must develop a continuous representation for $P_1$, which is larger in size and supports more initial conditions than $P_0$.

\section{Interpolant-mapping method}\label{sec.interpolant_mapping}
In the previous section we introduced a refinement method to produce a suitable discretization of the exact manifold to any desired resolution and extent. Despite our control over the calls to the map function and a precise definition of the refinement regions, the resulting procedure can be computationally intensive (but less intensive that a non-refining procedure). To reduce the computational cost we can move to an approximation paradigm, where the refinement of segment $P_n$ is based on approximated initial conditions in segment $P_{n-1}$, instead of initial conditions in $P_0$.

In this section we concentrate on producing an approximated continuous representation for $P_{n+1}$, based on an approximated representation of $P_n$. This is formally closer to the original composition presented in (\ref{eq.unst_man_union}), but, in practice, we use the refinement methods presented in the previous section to define an iterative scheme that allow us to produce reliable interpolants from a previous segment in the manifold.

Assume that we have a planar parametric curve $\bar P_n(s): \mathbb{R}\rightarrow \mathbb{R}^2$ for $s\in[s_l,s_r]$, that approximates the points in the $n$'th segment $P_n$ to a given precision $\varepsilon$. This is
\begin{equation}
 \forall s\in [s_l, s_r]\mbox{ , } \exists x\in P_n / |\bar P_n(s)-x| < \varepsilon.
\end{equation}
In other words the points in $\bar P_n(s)$ differ from those in $P_n$ by at most a distance $\varepsilon$.
To approximate the segment $P_{n+1}$ we can infer its behavior from an appropriate set of nodes
\begin{equation}
 x_{i} = T(\bar P_n(s_i)); i= 0,1,2,..., N,
\end{equation}
where $s_0$ and $s_N$ are the parameters corresponding to the end-points of the segment $P_{n+1}$. The set $s_0, s_1,..., s_N$ is chosen so that the curve between any pair $\{x_i, x_{i+1}\}$ can not differ much from a straight line. The procedure to do this is analogous to that presented in Fig.~\ref{fig.refinement_procedure}, where a new node between $x_i$ and $x_{i+1}$ is added when the resolution criteria are not meet. The new point is obtained by parametric bisection
\begin{equation}
 x'=T \left(\bar P_n \left(\frac{s_i+s_{i+1}}{2}\right)\right).
\end{equation}
After determining an appropriate set of nodes for the whole domain of $P_{n+1}$, the interpolant curve to approximate it is a piecewise vector function
\begin{equation}
 \bar P_{n+1}(s) = \left\{
 \begin{array}{ccl}
  f_0(\gamma_0, s) &,& s\in [s_0, s_1] \\
  \vdots &  & \\
  f_{N-1}(\gamma_{N-1}, s) &,& s\in [s_{N-1}, s_N] \\
 \end{array}
\right.,
\end{equation}
where $f_i:\mathbb{R}\rightarrow\mathbb{R}^2$ and $\gamma_i$'s are adjustable parameters that determine the local properties of the curve, and can be determined by imposing geometrical constrains on the interpolant pieces $f_i$.

\subsection{The interpolant model}
\begin{figure}[h]
 \centering
 \includegraphics[]{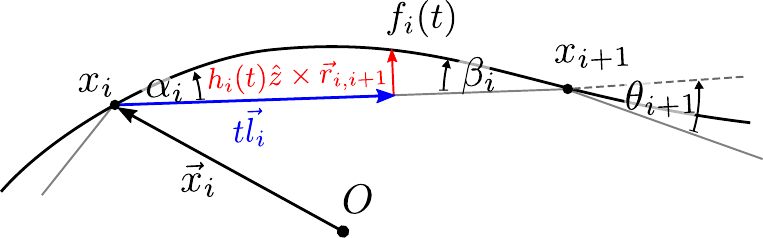}
 \caption{\label{fig.interpolant_vectors} The position along the model curve can be decomposed in three vectors. One defines the first node $x_i$, the second is the displacement along the chord joining two nodes and the third gives the normal displacement of the arc respect to the chord.}
\end{figure}

Consider the following expression for the position vector along interpolant of the $i'th$ arc
\begin{equation}\label{eq.curve_definition}
 \vec f_i(\gamma_i, s) = \vec x_i + t\vec l_i + h_i(\gamma_i, t)\hat z\times\vec l_i
\end{equation}
where $\vec x_i$ is the position vector of node $x_i$, $\vec l_i = \vec x_{i+1}-\vec x_i$, and $t\in[0,1]$ is the normalized position along the chord line joining $x_i$ and $x_{i+1}$ (see Fig.~\ref{fig.interpolant_vectors}). The function $h_i(\gamma_i, t):\mathbb{R}\rightarrow\mathbb R$ is proportional to the normal distance between the interpolant curve $f_i(s)$ and the chord $\overline{x_i, x_{i+1}}$, and satisfies $h_i(0) = h_i(1) = 0$, which guarantees that the curve passes through the nodes $x_i$ and $x_{i+1}$. Notice also that the L.H.S in (\ref{eq.curve_definition}) depends on the global parameters $s$ and the R.H.S depends on the normalized parameter $t$, so that there is a function $s = g(t)$ that allow us to make the transformation from the local parameter $t$ to the global one $s$.
The interesting part of this arc representation is that the bare geometrical aspects of the curve are contained in the first half of the expression $\vec x_i + t\vec l_i$, and the fine tunning is carried by the \emph{shape function} $h_i(\gamma_i, t)$, in the second part, which contains the error adjustable parameters $\gamma_i$. Notice also that a good choice of the partition nodes results in $h_i(\gamma_i,t) << 1$, everywhere in the interval, guaranteeing that the main geometry is well described by the nodes themselves and only small (but extremely important) details are described by $h_i(\gamma_i, t)$.

By definition, the arc representation (\ref{eq.curve_definition}) is continuous across the nodes $x_i$, but we also require the curve model to be differentiable everywhere. Since the interpolant is smooth inside the arcs we need to concentrate on the nodes and require
\begin{equation}
 \left.\frac{d\vec f}{ds}\right |_{s = s_i^-} = \left.\frac{d\vec f}{ds}\right |_{s=s_i^+},
\end{equation}
for all $s_i$ in the partitioning. In terms of the local representations of the interpolant this becomes
\begin{equation}
 \frac{dt}{ds}\left.\frac{d\vec f_{i-1}}{dt}\right |_{t=1} = \frac{dt'}{ds}\left.\frac{d\vec f_{i}}{dt'}\right |_{t'=0}
\end{equation}
where $t$ and $t'$ are the corresponding local parameters of the arcs $i-1$ and $i$. Because the global parameter $s$ has the same meaning across the node, the differential $ds$ is the same in both sides, and we are left with a simpler relation only involving the local parameters
\begin{equation}\label{eq.continuity}
 \left.\frac{d\vec f_{i-1}}{dt}\right |_{t=1} = \frac{dt'}{dt}\left.\frac{d\vec f_{i}}{dt'}\right |_{t'=0}.
\end{equation}
The relation $dt'/dt$ can be determined by the ratio of the projections of a line element $d\vec r$ along the secant lines $\overline{x_{i-1}x_i}$ and $\overline{x_ix_{i+1}}$, leading to
\begin{equation}
 \frac{dt'}{dt} = \frac{l_{i-1}}{l_i}\frac{\cos\alpha_{i}}{\cos\beta_{i-1}}
\end{equation}
where $\alpha_i$ and $\beta_{i-1}$ are the angles between the tangent line at $x_i$ and the secant lines. In other words, $\alpha_i$ and $\beta_i$ are the angles between the $i$'th arc and its corresponding chord line at the endpoints (see Fig.~\ref{fig.interpolant_vectors}). Writing $\beta_{i-1}$ as the difference between $\alpha_i$ and the inter-secant angle $\theta_i$, we obtain a consistent equation for the differentiability at the node $i$

\begin{equation}\label{eq.continuity}
 \left.\frac{d\vec f_{i-1}}{dt}\right |_{t=1} = \frac{l_{i-1}}{l_i}\frac{\cos\alpha_{i}}{\cos(\alpha_i-\theta_i)}\left.\frac{d\vec f_{i}}{dt'}\right |_{t'=0},
\end{equation}
Where the angles $\theta_i$ can be determined from the position of the nodes alone. For a suitable choice of the \emph{shape function} $h_i(\gamma_i, t)$, the chord-tangent angles $\alpha_i$ determine the overall behavior of the model curve and can be considered as the fundamental parameters of our interpolant.

Using the explicit arc representation in (\ref{eq.curve_definition}) and the differentiability condition (\ref{eq.continuity}) we can relate the the angles $\alpha_i$ with the derivatives of the shape function at the end-points

\begin{equation}\label{eq.endpoint_derivatives}
 h_i'(0) = \tan\alpha_i \mbox{ , } h_i'(1) = \tan (\alpha_{i+1}-\theta_{i+1}).
\end{equation}

\subsection{The simplest shape function}\label{sec.simplest_shape}
Requiring the shape function $h(\gamma_i,t)$ to be the lowest order polynomial, vanishing for $t=0$ and $t=1$  and allowing for different left and right derivatives we obtain a cubic function as the simplest expression for the shape function.
\begin{equation}\label{eq.cubic_shape_function}
 h_i(t) = a_it(1-t)^2 - b_i t^2(1-t),
\end{equation}
where $a_i = h'_i(0)$ and $b_i = h'_i(1)$ are the end-point derivatives, and $b_i$ can in turn be related to the parameters of the next shape function using the conditions obtained from differentiability in (\ref{eq.endpoint_derivatives}),
\begin{equation}\label{eq.relation_a-b}
 b_i = \frac{a_{i+1}-m_{i+1}}{1+a_{i+1} m_{i+1}},
\end{equation}
where $m_{i+1} = \tan\theta_{i+1}$. Given that the inter-chord angles $\theta_i$ are the resulting features of the discretization procedure, they are not adjustable parameters and are kept fixed. This leave us with the set $\vec a = \{a_0, a_1,..., a_{N}\}$ which can, in principle, be adjusted to obtain an interpolant $\bar P_{n+1}$ which is closest to the primary segment $P_{n+1}$. However, to adjust this set an error functional must be minimized, and this requires obtaining more information about the curve being modeled. For instance we can calculate a set of intermediate points $y_i = T(\bar P_n (s'_i))$ where $s'_i\in (s_i, s_{i+1})$ and minimize the functional
\begin{equation}
 \varepsilon (\vec a) = \sum_{i=0}^{N-1} |\vec f_i(a_i, a_{i+1}) - \vec y_i|,
\end{equation}
by performing subsequent variations $\delta \vec a$ obtained from a Levenberg-Marquardt procedure \cite{marquardt1963}. As expected, this results in an interpolant that passes through every node and is closest to the intermediate points $\vec y_i$. However, there can be an arbitrary distance between the interpolant and the primary segment $\bar P_{n+1}$ in every other location. In fact it was observed that an appropriate guess of the set $\{a_0, a_1, ..., a_N\}$ performed better at arbitrary locations than the numerically optimized set, even when additional optimization constrains were applied, like requiring the interpolant arc-length to be small or limiting the domain of variation of the $\alpha_i$'s.

A suitable guess consists in requiring the tangent line on each node to bisect the inter-secant angle. In other words $\alpha_i = \theta_i/2$, resulting the the following choice of parameters
\begin{equation}
 a_i = \frac{\sqrt{1+m_i^2}-1}{m_i} \mbox{ , }b_i = -a_{i+1}.
\end{equation}
with $m_i=\tan\theta_i$. This results in a very smooth curve passing through all the nodes, but as discussed before, the success of this approximation relies in leaving a minimum amount of information on the interpolant. Most of the information must be contained in the node positions, and the interpolant between two nodes must differ very little from a straight line, i.e. the values of $a_i$ must be very small. Notice that this is automatically satisfied for a good set of nodes because all inter-secant angles satisfy $\theta_i<\theta_c$ with $\theta_c <<1$.

We can estimate the maximum distance from the interpolant curve to the corresponding secant line in terms of the resolution parameters $l_c$ and $\theta_c$. Provided that $\theta_c<<1$ we have
\begin{equation}
 H_{c} = l_c h_{c} \approx \frac{\theta_c l_c}{8}.
\end{equation}
where $\theta_c$ is measured in radians and $h_c = h(0.5)$ for $a=b=\tan\theta_c$. This give us a baseline to measure the interpolation error, which must be small in units of $H_c$.

\subsection{Comparison to the exact calculation}

\begin{figure}[h]
 \centering
 \includegraphics[]{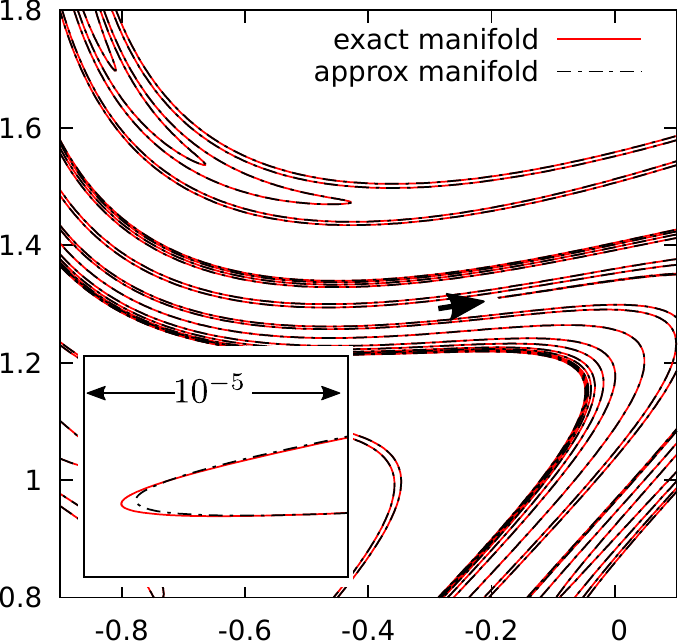}
 \caption{\label{fig.unst_man_compare} Comparison between $26$ primary segments of the exact and approximated unstable manifolds of the Chirikov-Taylor map for a region of phase space. To depict any difference between the manifolds we zoomed into an extreme corner, where differences below $10^{-6}$ develop.}
\end{figure}

In Fig.~\ref{fig.unst_man_compare} we compare the approximated manifold and the \emph{exact} one, in a $1\times 1$ region of the phase space for the Chirikov-Taylor map with $k=1.5$. For this comparison we traced $26$ primary segments with $d_c=0.01$ and {$\theta_c=3.0$\textdegree} for both the exact and approximated calculation. This gives us a distance baseline of $H_{c}\approx 6.5\times10^{-5}$, so that the local interpolation error must satisfy $\varepsilon<<H_{c}$. In the scale of Fig.~\ref{fig.unst_man_compare}, differences between the manifolds are not observable, and in fact they are not observable in any low or moderate-curvature region, because the distance between consecutive nodes of the manifold is much larger that the separation between the curves. To depict any difference we must look into very sharp corners, where any tiny amount of compression along the manifold amplifies the differences between the exact and the approximated manifold. 
 
\begin{figure}[h]
 \centering
 \includegraphics[width = 0.5\textwidth]{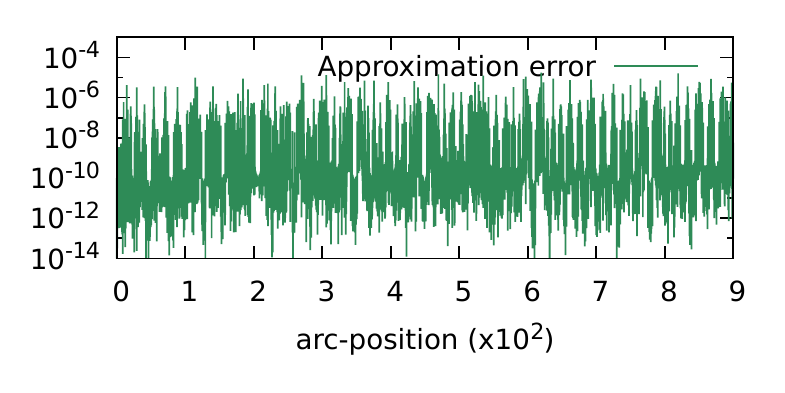}
 \caption{\label{fig.approx_error} Distance between the nodes of the exact manifold and their location from the manifold interpolant as a function of the arc-position along the manifold. The distance fluctuates widely but remains bounded below $10^{-5}$.}
\end{figure}

To have a better perspective of the approximated manifold error, we can measure the distance between an exact node and its expected position from the manifold interpolant. This can be done by finding the normal projection of the exact node along the secant line between two nodes of the approximated manifold. This give us the value of $t$ in (\ref{eq.curve_definition}) and we can calculate the corresponding interpolant position.

In Fig.~\ref{fig.approx_error} we show the distance between the  exact nodes and their corresponding estimation by the interpolants that compose the approximated manifold. Because the first segment of the approximated and exact methods are the same, we expect smaller errors for the first segments. However the correlation between the nodes of the exact calculation and approximated method is quickly lost because of the difference in the construction methods, then the distance from the exact nodes to their estimated positions give us a good estimate of the local error. An important feature of Fig.~\ref{fig.approx_error} is that the interpolant error does not grow along the manifold as a consequence of the approximation of new nodes based on a curve passing through previously approximated nodes. This indicates that the approximation method is stable, i.e. small errors introduced by the interpolation of new nodes get damped due to the compression perpendicular to the manifold.

\begin{figure}[h]
 \centering
 \includegraphics[width = 0.5\textwidth]{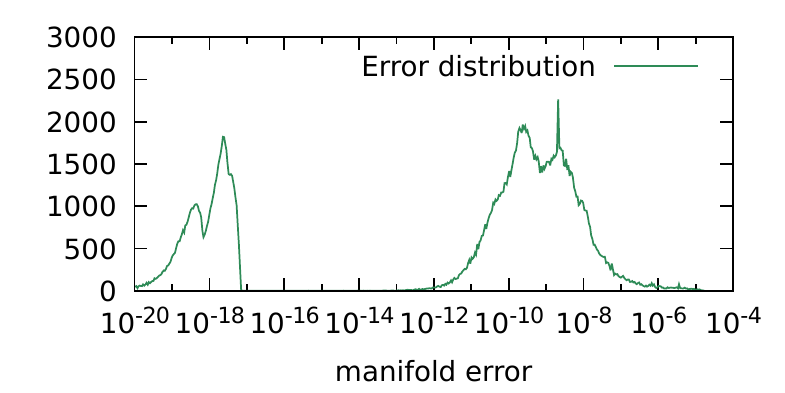}
 \caption{\label{fig.histogram_error} Statistical distribution of the distances in Fig.~\ref{fig.approx_error}. For a random position in the manifold there is a high probability of obtaining errors near $10^{-9}$ or $10^{-18}$ and the probability of values near the baseline $\sim 10^{-4}$ is significantly small.}
\end{figure}

For this example the manifold error is quite small, as can be seen in Fig.~\ref{fig.histogram_error}, where the statistical distribution of distances of Fig.~\ref{fig.approx_error} is shown. The error of the interpolants that compound the approximated manifold concentrate around $10^{-9}$ and $10^{-18}$, and error values close to the baseline $H_{max}\approx 6.5\times 10^{-5}$, show a very small probability. Remarkably, these results were obtained after setting the adjustable parameters $\{\alpha_0, \alpha_1,...\}$ only by geometrical requirements.

The resolution criteria led to $10^6$ calls to the map function for the exact calculation and $10^5$ calls for the approximated one. This factor \emph{ten} reduction in the number of calculations is due to the seeding of new initial condition in the segment previous to the segment under refinement, instead of the \emph{zero}'th segment. The number of calculations involved in the approximated calculation can be approximated by
\begin{equation}
 N^*_c(n) = \sum_{n'=0}^n M_{n'} \propto  \sum_{n'=0}^n e^{\lambda n'},
\end{equation}
so that, for a large stretching rate $\lambda$, and large number of primary segments $n$, the number of calls to the map approaches
\begin{equation}
 N_c^*(n) \propto \frac{e^{\lambda n}}{1-e^{-\lambda}},
\end{equation}
then, for a finite number of segments, the ratio to the number of calls for the exact calculations satisfy
\begin{equation}
 \frac {N_c^*}{N_c} \lesssim \frac{\kappa}{n},
\end{equation}
where $\kappa = (1-e^{-\lambda})^{-1} > 1$. This rapidly decreasing function of the number of segments $n$ guarantees a sustained growth in efficiency as the number of primary segments increases.

\section{An example application}\label{sec.example_application}
Now that we have illustrated the approximated manifold calculation for an explicit map, we move to a situation where the computational efficiency is more relevant, a continuous time dynamical system. Consider for instance the conservative Duffing oscillator, with Hamiltonian function
\begin{equation}
 H(q, p, t) = p^2/4 - 2q^2 + q^4 + \epsilon q \cos (\omega t),
\end{equation}
where $q$ and $p$ are conjugate variables and satisfy the Hamilton equations. Physically, $q$ and $p$ are the position and momentum of a particle in a double-well potential subjected to a periodic force field. In the unforced situation, particles with energy $E<0$ are restricted to one side of the well, but in the forced situation the external field enables the transition from one well to the other for energies moderately below \emph{zero}. This problem has interest in modeling the increase of the transition rate over potential barriers induced by external monochromatic fields~\cite{reichl1984}. Moreover, the corresponding deterministic diffusion in the chaotic region competes with quantum-mechanical tunneling between wells~\cite{dittrich1995, davis1986}.

In the subject of magnetically confined plasmas, this problem is topologically equivalent to that of a diverted single-null plasma~\cite{stangeby2000}, where the magnetic field lines confined to the plasma domain in an axisymmetric situation become wandering lines leaving the plasma when non-axisymmetric magnetic perturbations enter in consideration~\cite{da-silva2002,evans2002}. The use of resonant magnetic perturbations to break the magnetic invariants in the separatrix region has important applications in the control of plasma edge instabilities and plays an important role in modern tokamak operation.

Broadly, the potential barrier penetration can be understood through the formation of a chaotic layer around the separatrix of the well, which overcomes the potential barrier of the integrable case, allowing chaotic orbits to wander in both sides of phase space.
However, there are interesting situations where the transit between sides of the well is virtually suppressed for orbits in the chaotic layer during long extents of time, and manifold tracing provides relevant insight into the mechanisms involved in transport or lack of it.
\begin{figure}[h]
 \centering
 \includegraphics[width=0.45\textwidth]{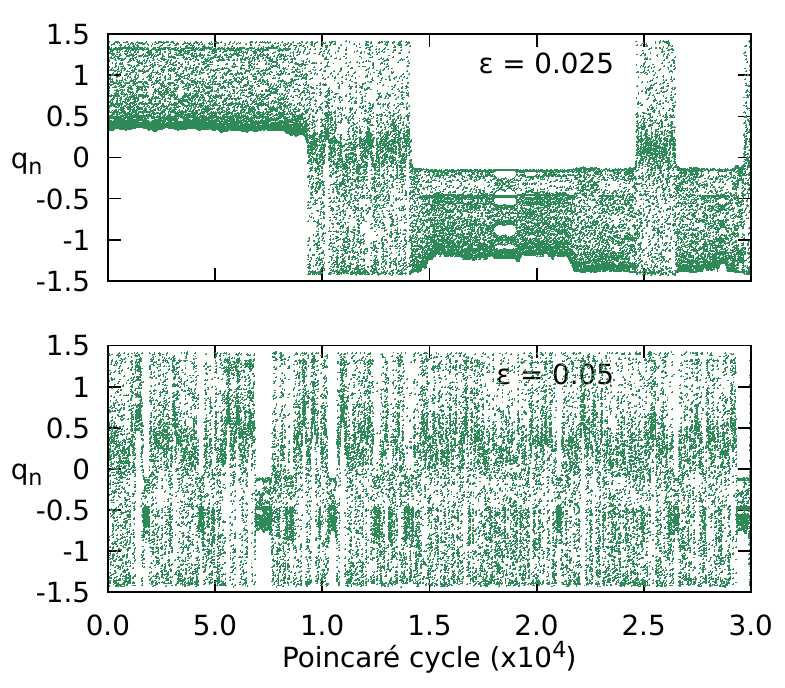}
 \caption {\label{fig.position_duffing} The particle's position from the stroboscopic map of the Duffing system with $\omega=1.5$ and two different amplitudes. Larger amplitudes lead to a predominantly global motion, while small ones cause spontaneous transitions between different the localized and global motions.}
\end{figure}

In Fig~\ref{fig.position_duffing} we show $30000$ cycles of the stroboscopic position of the Duffing system for an arbitrary orbit in the chaotic layer with different forcing amplitudes. For $\epsilon = 0.025$, the orbit spends a long time on each side of the well with shorter interludes of global motion. As we increase the amplitude, the times spent in local motions become shorter and global motion begins to dominate. For instance, at $\epsilon = 0.05$, periods of localized motion are very short compared to the global motion.

To understand these transitions in a geometrical basis we need to know which invariants control the motion in the chaotic region. From structural stability~\cite{guckenheimer1983}, it is expected for the perturbed system to contain a saddle $x^*_c$, which is a time-dependent version of the unperturbed saddle at $(0,0)$. The invariant manifolds of this saddle are responsible for driving the chaotic orbits from one side of the well to the other, but this is not the only period-\emph{one} saddle in the chaotic region. For $\omega=1.5$ there are two resonant tori near the separatrix of the double well, one for each side of the well. For $\epsilon=0.025$, they are destroyed by the perturbation creating period-\emph{one} K.A.M. islands surrounded by a thin chaotic layer driven by the invariant manifolds of a helical saddle $x^*_h$. The chaotic layers of the central saddle and the islands are in contact, but their interaction depends on the perturbation strength.

\begin{figure}[h]
 \centering
 \includegraphics[width=0.45\textwidth]{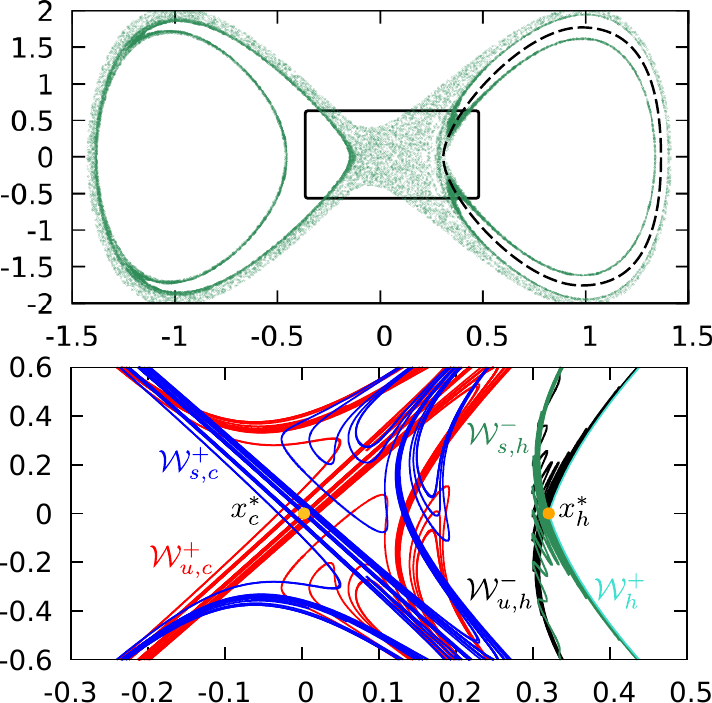}
 \caption{\label{fig.duffing_0p025} For $\omega = 1.5$ and $\epsilon=0.025$ the orbit transits between sides of the potential well spending long times near the helical saddles of period-\emph{one} K.A.M. islands. The resonant torus is sketched with a dashed line. The inset shows finite portions of the invariant manifolds of the central saddle $x_c^*$ and helical one $x_h^*$, exhibiting homoclinic intersections.}
\end{figure}
In Fig.~\ref{fig.duffing_0p025} we show a single orbit in the merged chaotic layers and an inset depicting the central and right period-\emph{one} saddles with some of their invariant manifolds. For $\epsilon=0.025$ a finite tracing of the invariants shows no entanglement between the manifolds of the central and the helical saddle. The lobes of the stable and unstable manifolds of $x^*_h$ develop around $x^*_h$, so that chaotic orbits near the helical saddle tend to remain so, leading to a slow diffusion from the chaotic layer around the island to the global layer responsible for transitions, as if they were only weakly connected. 

\begin{figure}[h]
 \centering
 \includegraphics[width=0.45\textwidth]{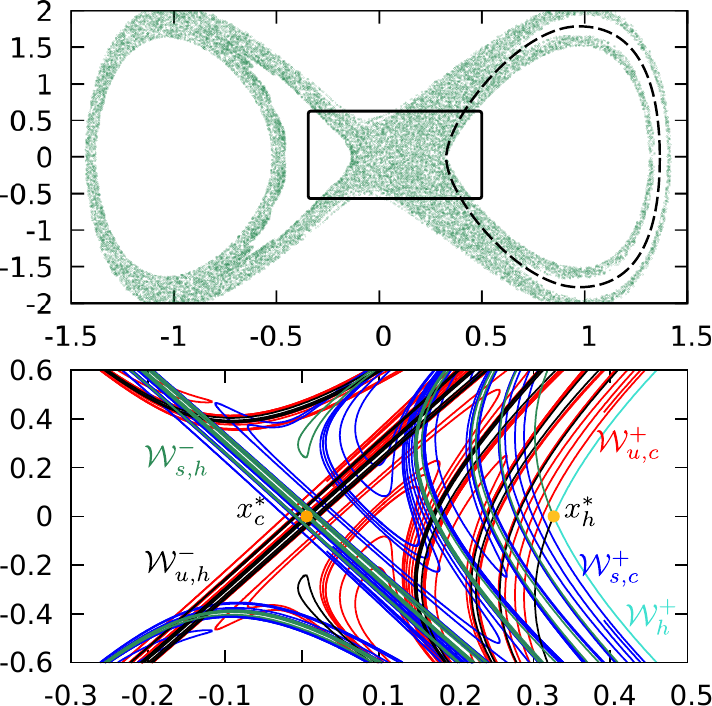}
 \caption{\label{fig.duffing_0p05} For $\omega = 1.5$ and $\epsilon=0.05$ the orbit wanders the chaotic region uniformly. The inset containing the central and helical saddles shows the development of heteroclinic connections between them, and a rather complex pattern of manifold intersections, so that diffusion in the chaotic region becomes more uniform.}
\end{figure}

When the forcing amplitude is increased the lobes of the stable and unstable manifolds of $x^*_h$ develop around $x^*_c$, and vice-versa, causing a heteroclinic connection between these saddles and a strong entanglement between their manifolds which results in a full mixing of the chaotic layers (Fig.~\ref{fig.duffing_0p05}). This transition from homoclinic to heteroclinic connection explains the transition from the alternating localized chaotic motion to the global chaotic dynamics, and leads to the  suppression of the local motion features of around the period-\emph{one} islands. Provided that there are transitions for the $\epsilon=0.025$ case, there must be some heteroclinic connection between the saddles, and such can be observed after a a rather extensive manifold tracing, however this is a second-order feature and the homoclinic connections determine the most relevant features of the weak forcing situation.

\section{Conclusions}\label{sec.conclusions}
In this paper we introduced a mapping-refinement approach to discretize the exact invariant manifolds of planar maps. This intensive method prevents redundant calculations and makes the best use of every new orbit resulting in a lower computational cost when compared to non-refining approaches. An intensive non-refinement technique with the same computational cost results in a non-smooth representation of the manifold with under-resolved and over-resolved portions.

Then we introduced an interpolant-mapping approach to approximate the invariant manifolds. Our interpolation method was based on the curve decomposition in bare a fine details, where the fine details were contained in a normal displacement function or~\emph{shape function}. This curve decomposition is new to our knowledge and does not involve any parametric optimization stage, which eases the method implementation.

With the approximation method the efficiency was greatly increased with a small precision penalty due to the interpolation the primary segments for seeding new orbits. Comparison between the exact and approximated manifold resulted in errors well below the determined baseline, even when the shape functions of the interpolant curve were chosen to be the simplest polynomial function with independent end-point derivatives. This leaves space for further improvement if more elaborate shape functions are used, or even parametric optimization techniques, with the cautions mentioned in Sect.~\ref{sec.simplest_shape}.

To illustrate the different manifold tracing routines we obtained the invariants for the Chirikov-Taylor map during the methods comparison, and used the approximation technique to study the chaotic transitions between local and global motion in the conservative Duffing oscillator, where it was evidenced the transition from homoclinic to heteroclinic connections between the central and helical saddles in the chaotic layer, in agreement with the observed intermittent dynamics.

\section*{Aknowledgements}
This work was developed with the financial support from the National Council for Scientific and Technological Development (CNPq, Brazil) grant No. 433671/2016-5, the São Paulo Research Foundation (FAPESP, Brazil) grants 2012/18073-1 and  2011/19296-1, and the US Department of Energy under DE-FC02-04ER54698 and DE-SC0012706.

\section*{References}
\bibliography{references}

\end{document}